\documentclass[aps,prd,preprintnumbers,superscriptaddress,nofootinbib,showpacs,twocolumn]{revtex4}%
\usepackage[dvipdfmx]{graphicx}
\usepackage{bm,latexsym,amsmath,amssymb,amsfonts,mathrsfs}
\usepackage{color}
\input{colordvi.tex}
\usepackage[dvipdfmx]{hyperref}
\usepackage{url}
\hypersetup{
    colorlinks=true,
    citecolor=cyan,
}
\newcommand*{\D}{{\rm d}}
\newcommand*{\mpl}{M_{\rm Pl}}
\begin{document}

\title{Healthy imperfect dark matter from effective theory of mimetic cosmological perturbations}

\author{Shin'ichi~Hirano}
\email[Email: ]{s.hirano"at"rikkyo.ac.jp}
\affiliation{Department of Physics, Rikkyo University, Toshima, Tokyo 171-8501, Japan}

\author{Sakine~Nishi}
\email[Email: ]{sakine"at"rikkyo.ac.jp}
\affiliation{Department of Physics, Rikkyo University, Toshima, Tokyo 171-8501, Japan}

\author{Tsutomu~Kobayashi}
\email[Email: ]{tsutomu"at"rikkyo.ac.jp}
\affiliation{Department of Physics, Rikkyo University, Toshima, Tokyo 171-8501, Japan}

\begin{abstract}
We study the stability of a recently proposed model of scalar-field matter called
mimetic dark matter or imperfect dark matter. It has been known that
mimetic matter with higher derivative terms
suffers from gradient instabilities in scalar perturbations.
To seek for an instability-free extension of imperfect dark matter,
we develop an effective theory of cosmological perturbations subject to
the constraint on the scalar field's kinetic term. This is done by using the
unifying framework of general scalar-tensor theories based on the ADM formalism.
We demonstrate that it is indeed possible to construct
a model of imperfect dark matter which is free from ghost and gradient
instabilities.
As a side remark, we also show that mimetic $F({\cal R})$ theory is plagued with
the Ostrogradsky instability.
\end{abstract}

\preprint{RUP-17-6}
\maketitle

\section{Introduction}

Standard cosmology based on Einstein's theory of general relativity
is very successful under the assumption that unknown components
called dark matter and dark energy dominate the energy density of
the Universe. The origins of those dark components are puzzles
in modern cosmology and particle physics, and
a number of scenarios have been developed and explored so far.

Recently, a novel interesting model of scalar-field matter
dubbed {\em mimetic dark matter}
was put forward~\cite{Chamseddine:2013kea}.
(For earlier work see
Refs.~\cite{Lim:2010yk,Gao:2010gj,Capozziello:2010uv},
and for a review see Ref.~\cite{Sebastiani:2016ras}.)
The mimetic scalar field is generated by
a singular limit of the general disformal transformation
where the transformation is not invertible~\cite{Deruelle:2014zza,Yuan:2015tta,Domenech:2015tca},
and its kinetic term is subject to the constraint
\begin{align}
g^{\mu\nu}\partial_\mu\phi\partial_\nu\phi =-1.
\end{align}
With this mimetic constraint the theory reproduces
the behavior of pressureless dust in Einstein gravity,
and thereby yields a candidate of dark matter.
In the original version of the mimetic scalar-field theory,
there is no
nontrivial dynamics for scalar-type fluctuations,
but by introducing the higher-derivative term
$(\Box\phi)^2$ the scalar degree of freedom can be promoted to
a dynamical field propagating with a
nonzero sound speed~\cite{Chamseddine:2014vna,Mirzagholi:2014ifa}.
The higher-derivative term modifies the fluid properties of mimetic dark matter,
and due to its imperfect nature it is called {\em imperfect dark matter}.
(For the Hamiltonian analysis of mimetic matter, see Refs.~\cite{Chaichian:2014qba,Malaeb:2014vua,Ali:2015ftw}.)
Since a nonzero sound speed affects the evolution of density perturbations
on small scales, imperfect dark matter could be relevant to
the missing-satellites problem and the core-cusp problem~\cite{Capela:2014xta}.
Essentially equivalent theories appear in
different theoretical settings such as
Ho\v{r}ava-Lifshitz gravity~\cite{Horava:2009uw,Ramazanov:2016xhp}, the Einstein-aether
theory~\cite{Jacobson:2000xp,Jacobson:2014mda}, and
non-commutative geometry~\cite{Chamseddine:2014nxa}.
Mimetic theories with more general higher derivative terms
can resolve black hole and cosmological
singularities~\cite{Chamseddine:2016uef,Chamseddine:2016ktu}.
The mimetic constraint can also be implemented in the general
second-order scalar-tensor theory
(the Horndeski theory~\cite{Horndeski:1974wa,Deffayet:2011gz,Kobayashi:2011nu}),
allowing for a variety of cosmological expansion
histories with fluctuations having a vanishing sound speed~\cite{Haghani:2015iva,Arroja:2015wpa,Rabochaya:2015haa,Arroja:2015yvd,Cognola:2016gjy}.
See also Refs.~\cite{Barvinsky:2013mea,Matsumoto:2015wja,Ramazanov:2015pha,Hammer:2015pcx,Liu:2017puc}
for further developments in mimetic dark matter.

Although mimetic dark matter (or, more generically, {\em mimetic gravity})
has thus received much attention, it is not free from problematic issues.
As it is anticipated from the fact that
mimetic gravity of~\cite{Chamseddine:2014vna,Mirzagholi:2014ifa}
can be reproduced as a certain limit of
Ho\v{r}ava-Lifshitz gravity~\cite{Horava:2009uw},
the two theories share the various aspects, some of which
could signal the sickness. For instance,
scalar perturbations exhibit gradient instabilities~\cite{Sotiriou:2009bx,Ijjas:2016pad}.
To what extent such gradient instabilities are dangerous depends on
the time scale on which perturbations grow,
but it is better if one could remove this potential danger from the theory
in the first place.
Another problem is an appearance of a caustic singularity~\cite{Babichev:2016jzg},
though it is a generic nature of scalar-tensor theories~\cite{Babichev:2016hys,Tanahashi:2017kgn}
rather than a problem specific to mimetic/Ho\v{r}ava-Lifshitz gravity,
and there exist some mechanisms to
avoid formation of the caustic~\cite{Mukohyama:2009tp,Babichev:2017lrx}.

In the present paper, we explore the way of resolving
one of the above fundamental problems, namely,
the gradient instability in
mimetic gravity. For this purpose, we consider a unifying framework of
general scalar-tensor theories~\cite{Gleyzes:2013ooa,Tsujikawa:2014mba,Gao:2014soa}
with the mimetic constraint,
based on which we develop an effective theory of cosmological perturbations
in mimetic theories.
This allows us to study the perturbation property of
mimetic gravity systematically at the level of the action rather than
at the level of the equations of motion.
It turns out that avoiding instabilities is not so difficult.
We present a concrete example of a simple extension of
mimetic gravity that is free from gradient as well as ghost instabilities.

This paper is organized as follows.
In the next section we give a short review on the mimetic dark matter model
of~\cite{Chamseddine:2013kea,Chamseddine:2014vna,Mirzagholi:2014ifa}
and its instability. We also argue briefly how one can remedy this gradient instability.
In Sec.~III we derive an effective theory of mimetic cosmological perturbations
from a general action of a mimetic scalar-tensor theory.
After presenting some unstable examples, we demonstrate that
a healthy extension of mimetic gravity is indeed possible.
We summarize our results and discuss future prospects in Sec.~IV.
As a side remark, in the appendix we show that the $F({\cal R})$ extension of mimetic gravity
exhibits Ostrogradsky instabilities and hence is not viable.

\section{Instabilities in mimetic gravity}

\subsection{Mimetic gravity}

We start with a brief review on mimetic gravity and
its cosmology~\cite{Chamseddine:2013kea,Chamseddine:2014vna,Mirzagholi:2014ifa}.
The action for mimetic gravity can be written in the form\footnote{In the present paper
we use the metric signature $(-,+,+,+)$,
which is different from the convention in
Refs~\cite{Chamseddine:2013kea,Chamseddine:2014vna,Mirzagholi:2014ifa}.}
\begin{align}
S&=\int\D^4 x\,{\cal L},
\\
\frac{{\cal L}}{\sqrt{-g}}&=\frac{{\cal R}}{2}-\lambda\left(g^{\mu\nu}\partial_\mu\phi\partial_\nu\phi+1\right)
 -V(\phi)+\frac{\alpha}{2}\left(\Box\phi\right)^2,\label{eq:original-Lagrangian}
\end{align}
where ${\cal R}$ is the Ricci scalar,
$\phi$ is the mimetic scalar field,
and $\lambda$ is the Lagrange multiplier enforcing
\begin{align}
g^{\mu\nu}\partial_\mu\phi\partial_\nu\phi+1=0.\label{eq:mimetic-constraint}
\end{align}
The parameter $\alpha$ can be a function of $\phi$,
but for simplicity we assume that it is a nonzero constant.
Here and hereafter we use the units $\mpl = 1$.

If $V(\phi)=\Lambda=\,$const, this theory is equivalent to
the IR limit of projectable Ho\v{r}ava-Lifshitz gravity~\cite{Horava:2009uw,Mukohyama:2009mz,Ramazanov:2016xhp}.
It is also equivalent to a particular class of the Einstein-aether theory
if the gradient of the mimetic scalar field is identified as
the aether vector field~\cite{Jacobson:2000xp,Jacobson:2014mda}.

The mimetic constraint~(\ref{eq:mimetic-constraint}) shows that
$\partial_\mu\phi$ must be a timelike vector.
It is therefore convenient to take the unitary gauge,
\begin{align}
\phi(t,\Vec{x}) = t,\label{eq:condition-phi=t}
\end{align}
and express the action in the Arnowitt-Deser-Misner (ADM) form in terms of
three-dimensional geometrical objects on constant time hypersurfaces,
i.e., the extrinsic and intrinsic curvature tensors, $K_{ij}$ and $R_{ij}$.
Such a method has been employed broadly in studies of
scalar-tensor theories of modified gravity~\cite{Gleyzes:2013ooa,Tsujikawa:2014mba,Gao:2014soa,Gleyzes:2014dya}.
The ADM decomposition of spacetime leads to the metric
\begin{align}
\D s^2=-N^2\D t^2+\gamma_{ij}\left(\D x^i+N^i\D t\right)
\left(\D x^j+N^j \D t\right),
\end{align}
where $N$ is the lapse function, $N^i$ is the shift vector,
and $\gamma_{ij}$ is the three-dimensional metric.
The extrinsic curvature is then given by
\begin{align}
K_{ij}:=\frac{1}{N}E_{ij},\quad
E_{ij}:=\frac{1}{2}\left(\dot \gamma_{ij}-D_iN_j-D_jN_i\right),
\end{align}
where a dot denotes the derivative with respect to $t$ and
$D_i$ is the covariant derivative induced by $\gamma_{ij}$.
The unit normal to constant time hypersurfaces is written as
$n_\mu=-\nabla_\mu \phi/\sqrt{-\nabla_\nu\phi\nabla^\nu\phi} = -N\delta^0_\mu$.
Noting that the trace of the extrinsic curvature is given by
$K = \nabla_\mu n^\mu$, we have
\begin{align}
\Box \phi = -\frac{K}{N}+\frac{\Xi}{N},
\end{align}
where
\begin{align}
\Xi:=\frac{\dot N}{N^2}-\frac{N^i\partial_iN}{N^2}.
\end{align}
We thus obtain the Lagrangian for mimetic gravity in the ADM form as
\begin{align}
\frac{{\cal L}}{\sqrt{\gamma }}&=
\frac{N}{2}\left(R+K_{ij}K^{ij}-K^2\right)+\lambda\left(\frac{1}{N}-N\right)
\notag \\ &\quad-NV(t) +\frac{\alpha}{2N}\left(K-\Xi\right)^2,
\label{eq:ADM-metirc-original}
\end{align}

Variation with respect to $\lambda$ gives the mimetic constraint,
\begin{align}
N(t,\Vec{x})=1.\label{eq:solution-N=1}
\end{align}
Note that we have not used the temporal gauge degree of freedom
to set $N=1$; rather it was already used to impose Eq.~(\ref{eq:condition-phi=t}).
As a result of solving the Euler-Lagrange equation we have obtained
Eq.~(\ref{eq:solution-N=1}).

Variation with respect to $N$ yields
\begin{align}
0&= \frac{1}{2}\left(R-E_{ij}E^{ij}+E^2\right)-2\lambda-V
\notag \\ &\quad
-\frac{3\alpha}{2}E^2+\frac{\alpha}{2}
\left[\dot E-D_i\left(N^iE\right)\right],\label{eq:delNeq}
\end{align}
where we substituted $N=1$ after taking the variation.
In contrast to the case of Einstein gravity,
this equation is {\em not} an initial value constraint.
Equation~(\ref{eq:delNeq}) fixes the Lagrange multiplier $\lambda$
in terms of the other variables, and hence is not necessary
for the purpose of determining the dynamics of the metric.

Varying Eq.~(\ref{eq:ADM-metirc-original}) with respect to $N_i$, we obtain
\begin{align}
D_j\pi^{ij}=0, \quad \pi^{ij}:=E^{ij}-(1-\alpha)\gamma^{ij} E.
\label{eq:momentum-const-original}
\end{align}
Finally, variation with respect to $\gamma^{ij}$ leads to the
evolution equation for the three-dimensional metric,
\begin{align}
\frac{1}{\sqrt{\gamma}}\frac{\partial}{\partial t}\left(\sqrt{\gamma}\pi^{kl}\right)
\gamma_{ik}\gamma_{jl}+\cdots = 0.\label{eq:evolution-original}
\end{align}
In deriving these equations we substituted $N=1$ after taking the variation.
One can use Eqs.~(\ref{eq:momentum-const-original}) and~(\ref{eq:evolution-original})
to determine the evolution of the dynamical variables in the unitary gauge.
Equivalently, one may start from the reduced Lagrangian,
\begin{align}
\frac{{\cal L}}{\sqrt{\gamma}} = \frac{1}{2}\left(R+E_{ij}E^{ij}-E^2\right)-V+\frac{\alpha}{2}E^2,
\label{eq:reduced-Lag}
\end{align}
to derive Eqs.~(\ref{eq:momentum-const-original}) and~(\ref{eq:evolution-original}).
The Lagrangian~(\ref{eq:reduced-Lag}) is much easier to handle.

\subsection{Cosmology in mimetic gravity}\label{sec:cosmology-in-mim}

Let us first study the evolution of the cosmological background,
for which $N_i=0$ and $\gamma_{ij}=a^2(t)\delta_{ij}$
on constant $\phi$ hypersurfaces.
The mimetic constraint enforces $N(t)=1$.
The evolution equation reads~\cite{Chamseddine:2014vna}
\begin{align}
3H^2+2\dot H = \frac{2}{2-3\alpha}V,\label{eq:a-EL}
\end{align}
where $H:=\dot a/a$ is the Hubble parameter.
This is the only equation we can use to determine the evolution of the
scale factor.
Given $V=V(\phi)=V(t)$, one can integrate Eq.~(\ref{eq:a-EL})
to obtain $H=H(t)$, which contains one integration constant.
This integration constant is left undetermined because
the Hamiltonian constraint is missing in mimetic gravity,
as is clear from the fact that Eq.~(\ref{eq:delNeq})
can only be used to fix $\lambda$.

Of particular interest is
the case with $V=\Lambda=\,$ const.
In this case Eq.~(\ref{eq:a-EL}) can be integrated to give
\begin{align}
\frac{3(2-3\alpha)}{2}H^2 = \Lambda+\frac{C}{a^3},
\end{align}
where $C$ is an integration constant.
This equation may be identified as the Friedmann equation
in the presence of the energy component mimicking dark matter, $C/a^3$.
This component is essentially equivalent to
dust-like matter as an integration constant found earlier in the context of
Ho\v{r}ava-Lifshitz gravity~\cite{Mukohyama:2009mz}.
When viewed as a fluid, this mimetic dark matter has
the imperfect character~\cite{Mirzagholi:2014ifa} with
its clustering behavior modified from the usual dust,
and hence it could have some impact on
the missing-satellites problem and the core-cusp problem~\cite{Capela:2014xta}.

\subsection{Cosmological perturbations in mimetic gravity}

Let us next consider cosmological perturbations in mimetic gravity.
In the unitary gauge,
the mimetic constraint imposes $N=1$ even away from the homogeneous background.
Therefore, it is sufficient to consider
\begin{align}
&N_i=\partial_i\chi,
\\
&\gamma_{ij}=a^2e^{2\zeta}
\left(e^{h}\right)_{ij}
=a^2e^{2\zeta}\left(\delta_{ij}+h_{ij}+\frac{1}{2}h_{ik}h^k_j+\cdots\right)
,
\label{eq:def:pert}
\end{align}
where $\chi$ and $\zeta$ are scalar perturbations and
$h_{ij}$ is a transverse and traceless tensor perturbation.
Plugging Eq.~(\ref{eq:def:pert}) to the reduced Lagrangian~(\ref{eq:reduced-Lag})
and expanding it to second order in perturbations, we
can derive the quadratic actions for the tensor and scalar perturbations.

For the tensor sector we simply have
\begin{align}
S_h^{(2)}
=\frac{1}{8}\int\D^4x\, a^3 \left[\dot{h}_{ij}^{2}-\frac{(\partial_kh_{ij})^{2}}{a^{2}}\right].
\label{eq:tensor2}
\end{align}
Clearly, there is no pathology in the tensor sector of mimetic gravity,
as Eq.~(\ref{eq:tensor2}) is identical to the corresponding Lagrangian in Einstein gravity.

For the scalar sector we have
\begin{align}
S^{(2)} =\int\D^4x &\,a^3\biggl[
\frac{1}{a^2}(\partial\zeta)^2 -\frac{3}{2}(2-3\alpha)\dot\zeta^2
\notag \\ &\quad
+(2-3\alpha)\dot\zeta\frac{\partial^2\chi}{a^2}+\frac{\alpha}{2}
\left(\frac{\partial^2\chi}{a^2}\right)^2\biggr],\label{eq:L_S^2}
\end{align}
where we used the background equation~(\ref{eq:a-EL}).
Varying $S^{(2)}$ with respect to $\chi$,
we obtain
\begin{align}
\alpha \frac{\partial^2\chi}{a^2}+(2-3\alpha)\dot \zeta=0.\label{eq:sol-chi}
\end{align}
This equation explains the reason why we have assumed that $\alpha\neq 0$:
if $\alpha = 0$ then the curvature perturbation $\zeta$ would be nondynamical.
Substituting this back into Eq.~(\ref{eq:L_S^2}), we arrive at
\begin{align}
S_\zeta^{(2)}=\int\D^4x\,a^3\left[\left(\frac{3\alpha -2}{\alpha}\right)\dot\zeta^2
-(-1) \frac{(\partial\zeta)^2}{a^2}\right].
\end{align}
From this quadratic Lagrangian we see that
ghost instabilities can be avoided for $(3\alpha -2)/\alpha > 0$.
However, it is obvious from the wrong sign in front of the $(\partial\zeta)^2$ term
that the scalar sector suffers from gradient instabilities.
This issue has already been known in the context of Ho\v{r}ava-Lifshitz
gravity~\cite{Horava:2009uw,Sotiriou:2009bx,Blas:2009yd},
and was reemphasized in Ref.~\cite{Ijjas:2016pad} in the context of mimetic cosmology.

\subsection{Curing mimetic gravity: A basic idea}\label{subsec:cure}

The gradient instability explained above arises
from the curvature term $+\sqrt{\gamma}R\sim +2(\partial\zeta)^2/a^2$.
In Einstein gravity, the sign of the $(\partial\zeta)^2$ term
is flipped when one removes the perturbation of the lapse function
by using the constraint equations.
However, this procedure is absent in mimetic gravity and
thus the wrong sign remains.

A possible way of curing this instability is to extend the theory
so that the quadratic Lagrangian for the scalar perturbations
contains the term $\partial^2\chi\partial^2\zeta$.
Then,
the Euler-Lagrange equation for $\chi$
i.e., Eq.~(\ref{eq:sol-chi}), would contain $\partial^2\zeta$.
Substituting $\partial^2\chi \sim \partial^2\zeta+\cdots$
back to the Lagrangian, we would have extra contributions of the form
$(\partial\zeta)^2$ and $(\partial^2\zeta)^2$
that improve the stability.
In the ADM language, such terms arise from
\begin{align}
RK,\quad R_{ij}K^{ij},
\end{align}
which, in the covariant language, correspond to the terms
\begin{align}
{\cal R}\Box\phi,\quad {\cal R}_{\mu\nu}\nabla^\mu\nabla^\nu\phi.
\end{align}
These terms appear in mimetic Horndeski gravity as
\begin{align}
\left({\cal R}_{\mu\nu}-\frac{1}{2}g_{\mu\nu}{\cal R}\right)\nabla^\mu\nabla^\nu \phi
\sim R_{ij}K^{ij}-\frac{1}{2}RK.
\end{align}
Unfortunately, in this case two $\partial^2\chi\partial^2\zeta$ terms
cancel out. It is therefore necessary to detune the coefficients of
the two terms.

This observation motivates us to study a general mimetic scalar-tensor theory
in the unitary gauge
whose action is given by
\begin{align}
S&=\frac{1}{2}\int \D^4x\sqrt{\gamma}\biggl[
N\left(R+K_{ij}K^{ij}-K^2\right)
\notag \\ &\quad\quad
+2\lambda\left(\frac{1}{N}-N\right)
+c_0(t,N)
\notag \\&\quad\quad
+c_1(t,N)K^2+c_2(t,N)RK+\cdots\biggr],
\label{eq:mimeticXG3}
\end{align}
namely, a spatially covariant theory~\cite{Gao:2014soa}
with the mimetic constraint.
In fact, the Lagrangian~(\ref{eq:mimeticXG3}) is redundant.
The mimetic constraint enforces $N(t,\Vec{x})=1$,
and also in this general theory the Euler-Lagrange equation $\delta S/\delta N=0$
is used only to fix the Lagrange multiplier $\lambda$.
Therefore, we may instead start from the reduced action
\begin{align}
S&=\frac{1}{2}\int\D^4x\sqrt{\gamma}\{
R+E_{ij}E^{ij}-[1-c_1(t)]E^2
\notag \\ &\quad\quad + c_0(t)+c_2(t)RE+\cdots
\},
\end{align}
as we did in usual mimetic gravity.
To study cosmology and stability in such gravitational theories,
in the next section, we construct an
effective theory of cosmological perturbations
for a general class of scalar-tensor theories with the mimetic constraint.

In the context of Ho\v{r}ava gravity,
several ideas have been proposed to remedy
the sickness of the theory~\cite{Blas:2009yd,Blas:2009qj,Blas:2010hb,Horava:2010zj}.
For example,
in~\cite{Blas:2009qj} the lapse function is allowed to be position dependent:
$N=N(t,\Vec{x})$.
In the context of mimetic gravity,
such generalization (in the unitary gauge) is possible only by
giving up the mimetic constraint, which is no longer what we call mimetic gravity.
With this generalization,
the scalar degree of freedom ceases to behave as a dark matter component.
In \cite{Horava:2010zj} the sick scalar mode is removed by imposing extra symmetry.
In contrast to those previous attempts, we require that
the lapse function remains subject to the mimetic constraint, $N=1$,
and no extra symmetry is imposed to remove the dangerous scalar mode,
but we introduce the coupling between the extrinsic and intrinsic curvature tensors.

\section{Effective theory of mimetic cosmological perturbations}

\subsection{General Lagrangian}

Following Refs.~\cite{Gleyzes:2013ooa,Tsujikawa:2014mba},
we consider a general action of the form
\begin{align}
S=\int\D^4x \sqrt{\gamma}
L(E,{\cal S},R,{\cal Y},{\cal Z};t),
\end{align}
where
\begin{align}
{\cal S}:=E_{ij}E^{ij},\quad {\cal Y}:=R_{ij}E^{ij},\quad {\cal Z}:=R_{ij}R^{ij}.
\end{align}
Note, however, that we set $N=1$ from the beginning,
which is different from the situation considered in~\cite{Gleyzes:2013ooa,Tsujikawa:2014mba}.
We do not include terms such as
$E_i^jE_j^kE_k^i$ and $E_i^jE_j^kR_k^i$ for simplicity,
but it turns out that the inclusion of them do
not change the essential result on the quadratic action
for cosmological perturbations.
Splitting $E_{ij}$ into the background and perturbation parts as
$E_i^j=H\delta_i^j+\delta E_i^j$, we expand $L$ as
\begin{align}
L&=
L_0+{\cal F}\delta E+(L_R+HL_{\cal Y}) R+L_{\cal S}\delta E_{ij}\delta E^{ij}
\notag \\ &\quad
+\frac{{\cal A}}{2}\delta E^2
+\left({\cal C}-\frac{L_{\cal Y}}{2}\right)
R\delta E+L_{\cal Y}R_{ij}\delta E^{ij}
\notag \\ &\quad
+L_{\cal Z}R_{ij}R^{ij}
+\frac{{\cal G}}{2}R^2+\cdots,\label{eq:Lexpanded}
\end{align}
where
\begin{align}
L_0&:=L(3H,3H^2,0,0,0;t),
\\
{\cal F}&:=L_E+2HL_{\cal S},
\\
{\cal A}&:=L_{EE}+4HL_{E{\cal S}}+4H^2L_{{\cal SS}},
\\
{\cal C}&:=L_{ER}+2HL_{{\cal S}R}+\frac{L_{\cal Y}}{2}+HL_{E{\cal Y}}+2H^2L_{{\cal SY}},
\\
{\cal G}&:=L_{RR}+2HL_{R{\cal Y}}+H^2L_{{\cal YY}},
\end{align}
with $L_E:=\partial L/\partial E$, $L_{ER}:=\partial^2L/\partial E\partial R$, etc.
The ellipsis denotes higher order terms.
Notice that $R_{ij}$ itself is a perturbative quantity.

From the background part of the action we obtain
\begin{align}
{\cal P}(H,\dot H;t):=\frac{\D {\cal F}}{\D t}+3H{\cal F}-L_0=0.
\end{align}
One can integrate this equation to find the background solution, $H=H(t)$.
Since we do not have the constraint corresponding to the
Friedmann equation, one integration constant remains undetermined in the solution.

Let us move to the perturbation part.
By substituting Eq.~(\ref{eq:def:pert}) to Eq.~(\ref{eq:Lexpanded})
it is straightforward to compute the quadratic action for the
tensor and scalar perturbations.

The action for the tensor perturbations is given by
\begin{align}
S_h^{(2)}=\int\D^4x\frac{a^3}{4}\left[L_{\cal S}\dot h_{ij}^2
-\frac{{\cal E}}{a^2}\left(\partial_k h_{ij}\right)^2+\frac{L_{\cal Z}}{a^4}
\left(\partial^2h_{ij}\right)^2
\right].
\end{align}
The stability conditions read $L_{\cal S}>0$, $L_{\cal Z}\le 0$, and
\begin{align}
{\cal E}:=L_R+\frac{1}{2a^3}\frac{\D}{\D t}\left(a^3 L_{\cal Y}\right)
\ge 0.
\end{align}
The action for the scalar perturbations is given by
\begin{widetext}
\begin{align}
S^{(2)}=\int\D^4x \,& a^3 \biggl\{
2\left[{\cal E}-\frac{3}{a}\frac{\D}{\D t}(a{\cal C})\right]\frac{(\partial\zeta)^2}{a^2}
+\left(\frac{9}{2}{\cal A}+3L_{\cal S}\right)\dot \zeta^2
+2\left(3L_{\cal Z}+4{\cal G}\right)\left(\frac{\partial^2\zeta}{a^2}\right)^2
\notag \\ &
+\frac{1}{2}\left({\cal A}+2L_{\cal S}\right)\left(\frac{\partial^2\chi}{a^2}\right)^2
-\left(3{\cal A}+2L_{\cal S}\right)\dot \zeta\frac{\partial^2\chi}{a^2}
+4{\cal C}\frac{\partial^2\zeta\partial^2\chi}{a^4}
\biggr\}.\label{eq:qaction1}
\end{align}
\end{widetext}
The Euler-Lagrange equation $\delta S^{(2)}/\delta \chi = 0$ implies
\begin{align}
\frac{\partial^2\chi}{a^2}
=\left(\frac{3{\cal A}+2L_{\cal S}}{{\cal A}+2L_{\cal S}}\right)\dot \zeta
-\frac{4{\cal C}}{{\cal A}+2L_{\cal S}}
\frac{\partial^2\zeta}{a^2},
\end{align}
where we assumed that
\begin{align}
{\cal A}+2L_{\cal S}\neq 0.
\end{align}
Using this one can remove $\chi$ from the action~(\ref{eq:qaction1})
to get
\begin{align}
S_\zeta^{(2)}&=\int\D^4x\,a^3\left[
q_1\dot\zeta^2-q_2\frac{(\partial\zeta)^2}{a^2}-q_3\frac{(\partial^2\zeta)^4}{a^4}
\right],
\end{align}
where
\begin{align}
q_1&:=2\left(\frac{3{\cal A}+2L_{\cal S}}{{\cal A}+2L_{\cal S}}\right)L_{\cal S},
\\
q_2&:=-2{\cal E}+\frac{8}{a}\frac{\D}{\D t}\left(
\frac{a{\cal C}L_{\cal S}}{{\cal A}+2L_{\cal S}}
\right),
\\
q_3&:=-2\left(3L_{\cal Z}+4{\cal G}\right)+\frac{8{\cal C}^2}{{\cal A}+2L_{\cal S}}.
\end{align}
To ensure the stability it is required that
\begin{align}
q_1>0, \quad q_2\ge 0, \quad q_3\ge 0.
\end{align}
As it follows from the stability of the tensor sector that ${\cal E}>0$,
the presence of nonzero ${\cal C}$ is crucial for the
stability of the scalar sector.

Let us comment on the special case where
${\cal A}+2L_{\cal S}\equiv 0$.
This occurs
for instance in mimetic Horndeski
gravity~\cite{Haghani:2015iva,Arroja:2015wpa,Rabochaya:2015haa,Arroja:2015yvd},
whose Lagrangian is of the form $L\sim A_3(t)E+ A_4(t)(E^2-{\cal S})+\cdots$.
This case is analogous to
the $\alpha = 0$ limit of the simplest mimetic theory presented
in the previous section: we always have $\zeta=\,$const and
hence the curvature perturbation is nondynamical.
This result is consistent with the analysis of~\cite{Arroja:2015yvd}.

\subsection{Unstable examples}

We have argued that without the ${\cal C}$ term
mimetic scalar-tensor theories inevitably suffer from gradient instabilities.
Let us present two concrete examples that are more general than~(\ref{eq:original-Lagrangian})
but still are {\em not} healthy.

The first example is mimetic gravity generalized
to include higher derivatives in the form of a general function $f(\Box\phi)$.
This class of theories allows us to resolve
the cosmological and black hole singularities~\cite{Chamseddine:2016uef,Chamseddine:2016ktu}
as well as to reproduce the dynamics of loop quantum cosmology~\cite{Liu:2017puc}.
(For the Hamiltonian analysis of $f(\Box\phi)$ mimetic gravity,
see Ref.~\cite{Kluson:2017iem}.)
The Lagrangian is given by~\cite{Chamseddine:2016uef,Chamseddine:2016ktu}
\begin{align}
\frac{{\cal L}}{\sqrt{-g}}=\frac{{\cal R}}{2}-\lambda\left(g^{\mu\nu}\partial_\mu\phi
\partial_\nu\phi+1\right)
-V(\phi)+f(\Box\phi).\label{eq:fbpmodel}
\end{align}
In the ADM language, this Lagrangian corresponds to
\begin{align}
L=\frac{1}{2}\left(R+{\cal S}-E^2\right)-V(t)+f(E),
\end{align}
for which we have
\begin{align}
{\cal A}=-1+f_{EE},\quad
L_{\cal S}=L_R=\frac{1}{2},\quad
{\cal C}=0,
\end{align}
with $f_{EE}:=\partial^2 f/\partial E^2$. This implies that
\begin{align}
q_1=3-\frac{2}{f_{EE}},\quad q_2=-1<0,
\end{align}
i.e., the scenarios based on the Lagrangian of the form~(\ref{eq:fbpmodel})
are plagued with gradient instabilities.
This instability was pointed out for the first time in Ref.~\cite{Firouzjahi:2017txv}.
Our result consistently reproduces that of~\cite{Firouzjahi:2017txv}.

The second example is
the Lagrangian of the form
\begin{align}
L=B_4(t)R+B_5(t)\left({\cal Y}-\frac{1}{2}ER\right)+
L'(E,{\cal S};t).\label{eq:unstable2}
\end{align}
Also in this case we have ${\cal C}=0$, and hence
the two conditions
${\cal E}>0$ and $q_2=-2{\cal E}>0$
are not compatible.
This confirms the argument in Sec.~\ref{subsec:cure}.

The Lagrangian~(\ref{eq:unstable2}) typically appears
in the case of
mimetic Horndeski gravity~\cite{Haghani:2015iva,Arroja:2015wpa,Rabochaya:2015haa,Arroja:2015yvd}.
However, as noted above, the condition ${\cal A}+2L_{\cal S}\neq 0$
is not satisfied for the Horndeski terms.
If the $L'(E,{\cal S};t)$ part is detuned away from the
Horndeski form, $\zeta$ is dynamical and unstable.

\subsection{A healthy extension of imperfect dark matter}

To demonstrate that it is indeed possible to construct
a stable cosmological model subject to the mimetic constraint,
let us consider a following simple extension of
imperfect dark matter~\cite{Chamseddine:2014vna,Mirzagholi:2014ifa},
\begin{align}
L &=\frac{1}{2}\left[R+{\cal S}-(1-\alpha)E^2\right]
 +\beta(t) RE +\frac{\beta^2}{2\alpha} R^2,\label{eq:Lag-minimal}
\end{align}
where we assume for simplicity that $\alpha$ is a constant,
but $\beta$ is a time-dependent function.

Since $R=0$ on the homogeneous background,
the background equation remains the same as presented in
Sec.~\ref{sec:cosmology-in-mim} (with $V(t)=0$):
\begin{align}
3H^2+2\dot H = 0.
\end{align}
Thus, we consider a universe dominated by mimetic dark matter,
\begin{align}
3H^2=\frac{C}{a^3}
\quad
\Rightarrow
\quad
H=\frac{2}{3t}.
\end{align}

For the Lagrangian~(\ref{eq:Lag-minimal}) we have
\begin{align}
L_{\cal S}=\frac{1}{2}, \quad {\cal E}=L_R=\frac{1}{2}+3\beta H,\quad L_{\cal Z}=0,
\end{align}
and
\begin{align}
q_1 = \frac{3\alpha-2}{\alpha},
\quad
q_2=\frac{4\dot\beta}{\alpha}-\left(1+2q_1 \beta H\right),
\quad
q_3=0.
\end{align}
Since $L_{\cal Z}=q_3\equiv 0$,
we do not need to care about the higher spatial derivative terms.
Note that the coefficient in front of the $R^2$ term ($\beta^2/2\alpha$)
is tuned so that $q_3=0$. If one would instead want to have a
higher spatial derivative term
$\partial^4\zeta$
that could affect the perturbation evolution
on small scales,
one may introduce a slight deviation from this value.

To be more specific,
suppose that
\begin{align}
\alpha = -\varepsilon,\quad \beta= - \xi \varepsilon t=-\xi\varepsilon\phi,
\end{align}
with $0<\varepsilon\ll 1$ and $\xi={\cal O}(1)$. (Recall that we use the units $\mpl=1$.)
In this case, we have
$L_R=1/2-2\xi\varepsilon>0$.
The propagation speed of gravitational waves, $c_h$, is slightly subluminal
if $\xi>0$,
\begin{align}
c_h=\sqrt{\frac{L_R}{L_{\cal S}}}=1-{\cal O}(\varepsilon)<1.
\end{align}
Depending on the assumption about the origin of the high energy cosmic rays,
a lower bound on $c_h$ has been obtained
from gravitational Cherenkov radiation~\cite{Moore:2001bv,Elliott:2005va,Kimura:2011qn}.
In the case of the galactic origin, the constraint reads
\begin{align}
1-c_h<2\times 10^{-15},
\end{align}
which in turn sets the upper bound of $\varepsilon$.
For the scalar sector, we have
\begin{align}
q_1=\frac{2+3\varepsilon}{\varepsilon}>0,
\quad
q_2=\frac{20}{3}\xi -1+4\xi\varepsilon.
\end{align}
It can be seen that $q_2>0$ for $\xi\gtrsim 1$
and the sound speed, $c_s$, is very small,
\begin{align}
c_s=\sqrt{\frac{q_2}{q_1}}={\cal O}(\varepsilon^{1/2})\ll 1.
\end{align}

Thus, we see that a simple healthy extension of mimetic gravity
is indeed possible.
It is straightforward to promote the ADM Lagrangian~(\ref{eq:Lag-minimal})
to a manifestly covariant form, which will be reported, supplemented with
cosmological applications,
in a separate publication~\cite{Hirano:nextwork}.

\section{Summary and future prospects}

In this paper, we have developed a general theory of cosmological perturbations
in the unitary gauge
in scalar-tensor theories with the mimetic constraint,
focusing in particular on the gradient instability issue.
In the unitary gauge ($\phi(t,\Vec{x})=t$),
the mimetic constraint enforces $N(t,\Vec{x})=1$,
which greatly simplifies the analysis based on the ADM formalism.
We have presented a concrete stable example of a mimetic theory of gravity,
in which the mimetic scalar field plays the role of dark matter.
As the evolution of density perturbations on small scales is
modified in our new mimetic dark matter model,
it would be interesting to investigate the power spectrum,
which we leave for a future study.

For the purpose of stabilizing the scalar perturbations,
we have introduced the coupling between the curvature and the scalar degree of freedom.
This could modify the nature of a weak gravitational field
inside the solar system. For this reason the solar-system constraints
on our mimetic dark matter models should be studied.
It would also be interesting to address the problem of the formation of
a caustic singularity in mimetic dark matter.
We will come back to those issues in a future study.

\acknowledgements
We thank  Yuji Akita, Tomohiro Harada, Ryotaro Kase,
Shinji Mukohyama, Kazufumi Takahashi, Shuichiro Yokoyama, and Daisuke Yoshida
for useful comments and fruitful discussion.
This work was supported in part by the
MEXT-Supported Program for the Strategic Research Foundation at Private Universities, 2014-2017, the JSPS Research Fellowships for
Young Scientists No.~15J04044 (S.N.),
and by the JSPS Grants-in-Aid for Scientific Research No.~16H01102 and No.~16K17707 (T.K.).
\appendix

\section{Ostrogradsky instability in mimetic $F({\cal R})$ gravity}

In this appendix, we consider the $F({\cal R})$ extension of mimetic
gravity~\cite{Nojiri:2014zqa}
and show that the theory is plagued with the Ostrogradsky instability.

The action for mimetic $F({\cal R})$ gravity is given by~\cite{Nojiri:2014zqa}
\begin{align}
S=\int\D^4x\sqrt{-g}\left[\frac{F({\cal R})}{2}
-\lambda\left(g^{\mu\nu}\partial_\mu\phi\partial_\nu\phi+1\right)
 -V(\phi)\right].\label{mimeticFR}
\end{align}
Following the usual procedure in $F({\cal R})$ gravity,
we rewrite the first term by introducing an auxiliary scalar field $\varphi$ as
\begin{align}
F({\cal R})\;\to\;F'(\varphi)({\cal R}-\varphi)+F(\varphi),
\end{align}
where we assumed that $F''(\varphi)\neq 0$. Performing the conformal transformation
$\widetilde g_{\mu\nu}=e^{2\sigma}g_{\mu\nu}$
with $e^{2\sigma}=F'(\varphi)$, we obtain the action in the Einstein frame as
\begin{align}
S&=\int \D^4x \sqrt{-\widetilde{g}}
\biggl[
\frac{\widetilde{\cal R}}{2}-3\widetilde g^{\mu\nu}\partial_\mu\sigma\partial_\nu\sigma-U(\sigma)
\notag \\ &\quad
-e^{-2\sigma}\lambda\left(\tilde g^{\mu\nu}\partial_\mu\phi\partial_\nu\phi+e^{-2\sigma}\right)
-e^{-4\sigma}V(\phi)
\biggr],
\end{align}
where $U=\left(\varphi F'-F\right)/2(F')^2$.
We are thus lead to study a bi-scalar mimetic theory
whose action is of the form
\begin{align}
S&=\int\D^4x\sqrt{-g}\biggl\{\frac{{\cal R}}{2}
-3g^{\mu\nu}\partial_\mu\sigma\partial_\nu\sigma+P(\phi, \sigma)
\notag \\ &\quad
-\lambda\left[g^{\mu\nu}\partial_\mu\phi\partial_\nu\phi+2f(\sigma)\right]
\biggr\}.\label{ap:bi-scalar-S}
\end{align}
Here the redefinition of the Lagrange multiplier $\lambda$ was made.
The tildes were omitted for brevity from the Einstein frame metric.
To keep generality, we allow for general functions
$P(\phi,\sigma)$ and $f(\sigma)$.
In this sense mimetic $F({\cal R})$ gravity is a specific case
of the general theory described by~(\ref{ap:bi-scalar-S}).

It follows from $\delta S/\delta \lambda = 0$ that
\begin{align}
g^{\mu\nu}\partial_\mu\phi\partial_\nu\phi+2f(\sigma)=0,
\end{align}
which can be solved for $\sigma$ as
\begin{align}
\sigma = h(X), \quad X:=-\frac{1}{2}g^{\mu\nu}\partial_\mu\phi\partial_\nu\phi,\label{FR_solution}
\end{align}
where
$h$ is the inverse function of $f$.
Substituting this to the action~(\ref{ap:bi-scalar-S}), we obtain
\begin{align}
S=\int\D^4x\sqrt{-g}\left[
\frac{{\cal R}}{2}-3h_X^2g^{\mu\nu}\partial_\mu X \partial_\nu X +P(\phi, X)
\right].\label{ostaction1}
\end{align}
Now it is obvious that the field equation for $\phi$ is of fourth order,
implying the presence of the Ostrogradsky ghost.
Hence, mimetic $F({\cal R})$ gravity~\cite{Nojiri:2014zqa} is never viable.

Note that a singular disformal transformation
(i.e., imposing the mimetic constraint) does not commute with a regular
conformal transformation.
In other words, the frame in which the mimetic constraint is imposed could be crucial.
Indeed, transforming the $F({\cal R})$ action
to the scalar-tensor theory in the {\em Einstein frame} first and then
imposing the mimetic constraint do not lead to~(\ref{ostaction1}).
The resultant theory is original mimetic gravity plus another canonical scalar field,
which is free from the Ostrogradsky ghost.


\end{document}